\documentstyle[11pt,natb_209]{article} 

\def\dosingle#1::::{#1}  \def\dodouble#1::::{ } 

\dodouble \documentstyle[natb_209,doublespacing]{article} ::::
\input epsf1990.sty  
\twocolumn
\def\nice#1::::{#1}    \def\subm#1::::{}   

\newcommand\zzz[2]{#2} 

\def\SS{Sect.~}

\def\apj{ApJ}                 
\def\aap{A\&A}            
\def\mnras{MNRAS}
\def\araa{AnnRevA\&A}

\nice \def\mycaptionfont{\protect\footnotesize} ::::

\renewcommand\citep[1]{(\citealt{#1})}
\newcommand\citepf[1]{(\citealt*{#1})} 

\def\centreline{\centerline}
\def\.{{\cdot}} 
\def\gtapprox{\,\lower.6ex\hbox{$\buildrel >\over \sim$} \, }
\def\ltapprox{\,\lower.6ex\hbox{$\buildrel <\over \sim$} \, }
\def\propapprox{\,\lower.6ex\hbox{$\buildrel \propto\over \sim$} \, }

\def\arcs{\ifmmode {'' }\else $'' $\fi}     
\def\arcm{\ifmmode {' }\else $' $\fi}       
\def\deg{\ifmmode^\circ\else$^\circ$\fi}    
\def\ttimes{{\scriptstyle \times}}

\def\fr7{7$ \hskip -0.9ex \vrule height0.8ex width0.8ex depth-0.73ex
                                                     \hskip0.1ex$}
\def\frtoday{Le\space\number\day\space\ifcase\month\or
  janvier\or f\'evrier\or mars\or avril\or mai\or juin\or
  juillet\or ao\^ut\or septembre\or octobre\or novembre\or 
d\'ecembre\fi\space \number\year}

\newcommand\joref[5]{#1, #5, {#2, }{#3, } #4}  
 
\newcommand\epref[3]{#1, #3, #2}

\def\cqg{ClassQuantGra}   %

\def\hMpc{\mbox{$h^{-1}$ Mpc}}

\def\rinj{{r}_{\mbox{\rm \small inj}}}  

\def\exunit{{\bf e_x}}  
\def\eyunit{{\bf e_y}}  
  
\title{
A Counterexample to 
Claimed COBE Constraints on Compact Toroidal Universe Models} 

\author{Boudewijn F. Roukema\\
Inter-University Centre for Astronomy and Astrophysics \\
Post Bag 4, Ganeshkhind, Pune, 411 007, India\\ Email: boud@iucaa.ernet.in}

\date{\frtoday}

\begin{document}

\maketitle

\begin{abstract}
It has been suggested that if the Universe satisfies a 
flat, multiply connected, perturbed Friedmann-Lema\^{\i}tre model, then 
cosmic microwave background data from the COBE satellite implies that
the minimum size of the injectivity diameter
(shortest closed spatial geodesic) must be larger than about
two fifths of the horizon diameter.
To show that this claim is misleading, a simple
$T^2 \ttimes R$ universe model of injectivity diameter a quarter of this
size, i.e. a tenth of the horizon diameter, 
is shown to be consistent with COBE four year 
observational maps of the cosmic microwave background.
This is done using the identified circles principle. 

\end{abstract}

\dodouble \clearpage :::: 


\def\tbestlong{
\begin{table}
\caption{Sky positions of long axis for `good' $T^2$ hypotheses based on 
maximising the symmetry statistic $S_{T^2}$ 
[eq.~(\protect\ref{e-st2})]. Galactic
longitude, latitude and $S_{T^2}$ are listed. Adjacent pairs $(i,j)$ 
are excluded from the $S_{T^2}$ values shown, in order to decrease
noise from close pixels.
\label{t-bestlong}}
$$\begin{array}{c c c} \hline 
\rule{0ex}{2.5ex}
\mbox{$l^{{\sc II}}$} & \mbox{$b^{{\sc II}}$} & S_{T^2} \\ \hline 
       329\deg &     -42\deg  &  0.23 \\
      282\deg  &      32\deg  &  0.22 \\
       104\deg  &    -38\deg    & 0.21 \\
\hline
\end{array}$$
\end{table}
}  

\def\tbestshort{
\begin{table*}
\caption{
Axis positions of a 3-manifold candidate found close to a
long axis position listed in 
Table~\protect\ref{t-bestlong}, 
where the long axis ($Z_{T^2}$) 
is larger than the horizon diameter $2 R_H$ and 
$2\rinj \equiv 2R_H/10$ is the length of the two
short axes ($X_{T^2}$, $Y_{T^2}$). 
The KS probability of finding the observed temperature
differences given the 3-manifold as a null hypothesis is 
$P_{\mbox{\rm all}}$ or $P_{\mbox{\rm subs}}$, depending on whether
the full set of circles (where $N=150$ independent pairs are
assumed) or an evenly spaced
subset of $N=138$ pairs of circles, respectively, is used.
An ISW/systematic noise contribution of $x^2=0.3$ 
[eq.~(\protect\ref{e-xdefn})] is adopted. Statistics $\sigma,$ $d$ and
$S$ are defined in eqs~(\protect\ref{e-sigma}), 
(\protect\ref{e-dmean}) and (\protect\ref{e-corr}) respectively.  
\label{t-bestshort}}
$$\begin{array}{c c c c c c c c ccc} \hline 
 \multicolumn{2}{c}{\mbox{\rm long ($Z_{T_2}$)}} & 
\multicolumn{2}{c}{\mbox{\rm short ($X_{T_2}$)}} & 
\multicolumn{2}{c}{\mbox{\rm short ($Y_{T_2}$)}} & 
P(\mbox{\rm all}) & P(\mbox{\rm subs}) & 
\sigma & d & S \\
\mbox{$l^{{\sc II}}$} & \mbox{$b^{{\sc II}}$} & 
\mbox{$l^{{\sc II}}$} & \mbox{$b^{{\sc II}}$} & 
\mbox{$l^{{\sc II}}$} & \mbox{$b^{{\sc II}}$} \\ 
\hline 
280 & 37.5 & 184 & 8 & 264 & -51 & 0.40 & 0.19
& 1.61 & -0.006 & 0.21 \\
\hline
\end{array}$$
\end{table*}
}  

\def\tworst{
\begin{table*}
\caption{Axis positions, statistics and null hypothesis 
probabilities,  
as for Table~\ref{t-bestshort},
for a $T^2$, $2\rinj=2R_H/10$ 
model strongly rejected using the
identified circles principle, in spite of the presence of 
a strong ISW effect, with $x^2=0.6$. 
\label{t-worst}}
$$\begin{array}{c c c c c c c c ccc} \hline 
 \multicolumn{2}{c}{\mbox{\rm long ($Z_{T_2}$)}} & 
\multicolumn{2}{c}{\mbox{\rm short ($X_{T_2}$)}} & 
\multicolumn{2}{c}{\mbox{\rm short ($Y_{T_2}$)}} & 
P(\mbox{\rm all}) & P(\mbox{\rm subs}) & 
\sigma & d & S \\
\mbox{$l^{{\sc II}}$} & \mbox{$b^{{\sc II}}$} & 
\mbox{$l^{{\sc II}}$} & \mbox{$b^{{\sc II}}$} & 
\mbox{$l^{{\sc II}}$} & \mbox{$b^{{\sc II}}$} \\ 
\hline 
 191&   -57.5 &   325 &      -24 &      244 &       21 
& 0.002  & 0.001
& 1.7 & -0.25 & -0.02 \\
\hline
\end{array}$$
\end{table*}
}  

\def\fprobISW{
\begin{figure}
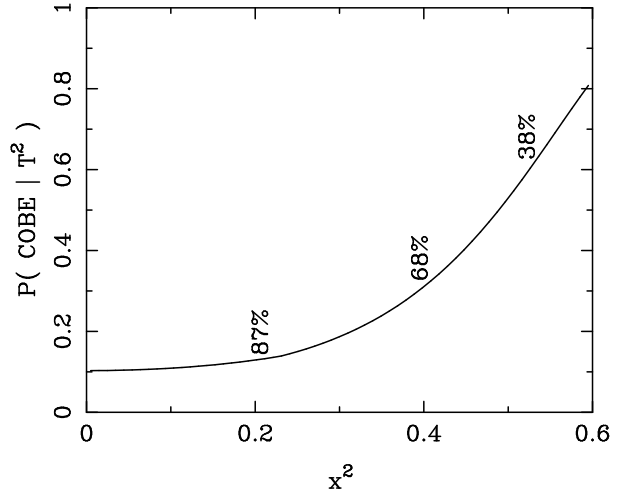

\centering 
\nice \centreline{\epsfxsize=8cm
\zzz{\epsfbox[42 34 470 382]{"`gunzip -c roukfig1.eps.gz"}}
{\epsfbox[42 34 470 382]{"roukfig1.eps"}}  } ::::
\caption[]{ \mycaptionfont
Probability, $P_{\mbox{\rm subs}}$ 
given the small $T^2$ model indicated in
Table~\protect\ref{t-bestshort} as a null hypothesis, that the
differences around matched circles in the COBE data are simply due to
random error, integrated Sachs-Wolfe contributions and other
systematic (or random) error, where $x^2$ 
[eq.~(\protect\ref{e-xdefn})] represents the
contribution of the latter two. Some significance levels for rejection of 
the hypothesis are labelled at appropriate points. A rejection of
95\% (which would correspond to a two--tailed Gaussian rejection at 
two standard deviations if the probability distribution were 
Gaussian) is not attained, even for $x^2=0$ (no ISW contribution),
where $1-P=92\%$.

}
\label{f-probISW}
\end{figure} 
} 

\def\fcircleCE{ 
\begin{figure*}
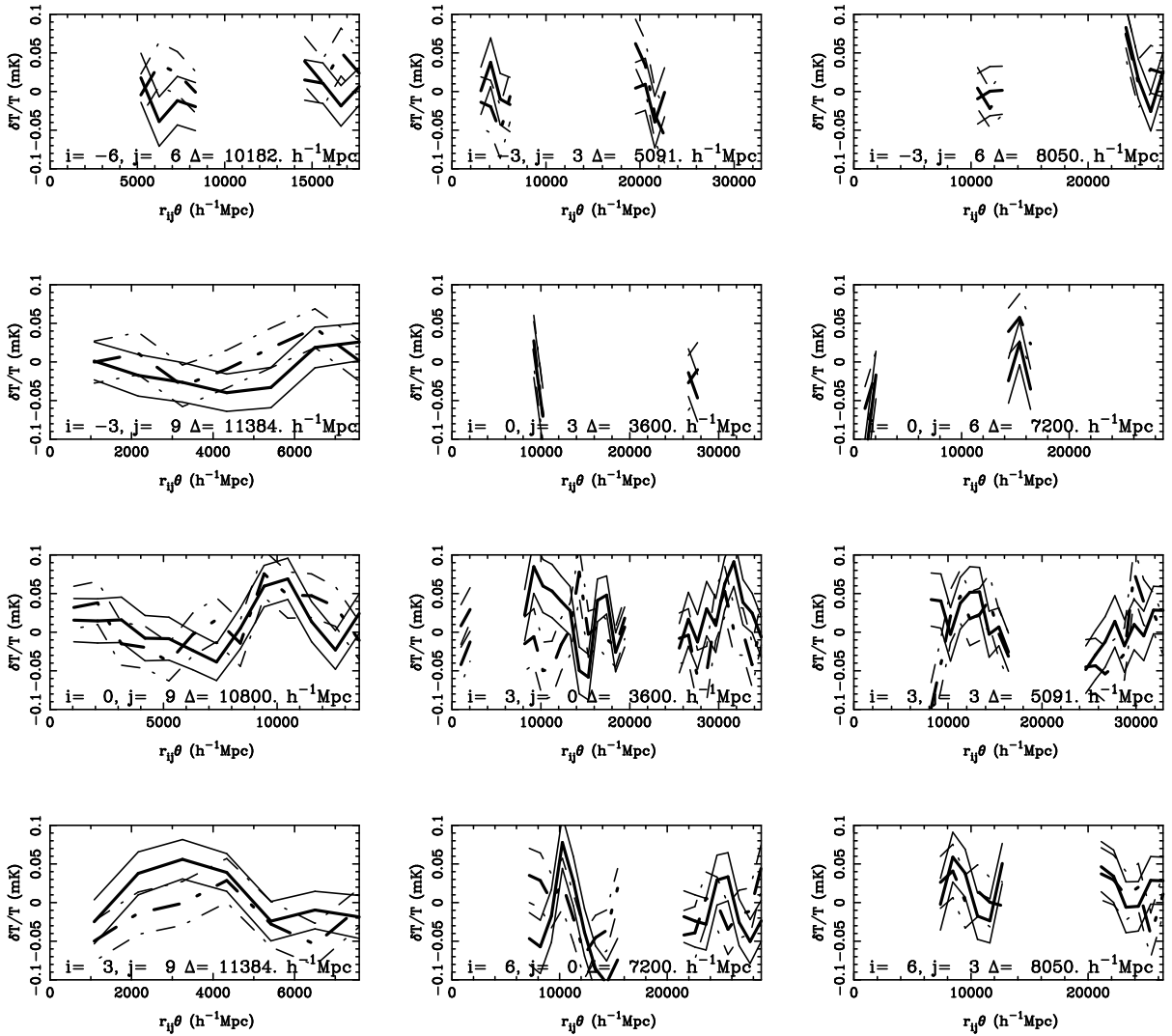

\centering 
\nice \centreline{\epsfxsize=16cm
\zzz{\epsfbox[40 280 581 761]{"`gunzip -c roukfig2.eps.gz"}}
{\epsfbox[40 280 581 761]{"roukfig2.eps"}}  } ::::
\caption[]{ \mycaptionfont
Temperature fluctuations in four year COBE DMR data 
around a subset of identified
circles in the CMB, in the covering
space for $\Omega_0 + \lambda_0=1$, for a $T^2$ universe
having $2\rinj=2R_H/10$ and oriented as indicated in 
Table~\protect\ref{t-bestshort}, shown against the distance
around each circle assuming $(\Omega_0=1,\lambda_0=0)$. 
Thick lines are $\delta T/T$ 
and thin lines are $\delta T/T$ $\pm \Delta (\delta T/T)$ uncertainties 
for $x=0.3,$ i.e. including a component for the ISW effect 
($\Omega_0 < 1, \lambda_0=1-\Omega_0$) and/or
unaccounted for systematic error. Solid and dot-dashed lines
distinguish the members of each circle pair.
The horizontal length of each panel is the circle circumference 
if $(\Omega_0=1,\lambda_0=0)$. 
Circles are labelled $(i,j)$,  where each circle lies in 
a plane halfway between the observer 
and her/his topological image at 
$i (2R_H/10) \exunit + j (2R_H/10) \eyunit$ 
and its matching circle is at $(-i,-j),$ 
where $\exunit$ and $\eyunit$ are unit vectors in the two short directions.
The distance between a circle and its match is
indicated here by $\Delta.$ Galactic latitudes with 
$|b^{\mbox{{\sc II}}}| < 20\deg$ are excluded.
}
\label{f-circleCE}
\end{figure*} 

} 

\def\fpolar{ 
\begin{figure*}
\centering 
\nice \centreline{\epsfxsize=14cm
\zzz{\epsfbox[55 111 787 484]{"`gunzip -c roukfig3.eps.gz"}} 
{\epsfbox[55 111 787 484]{"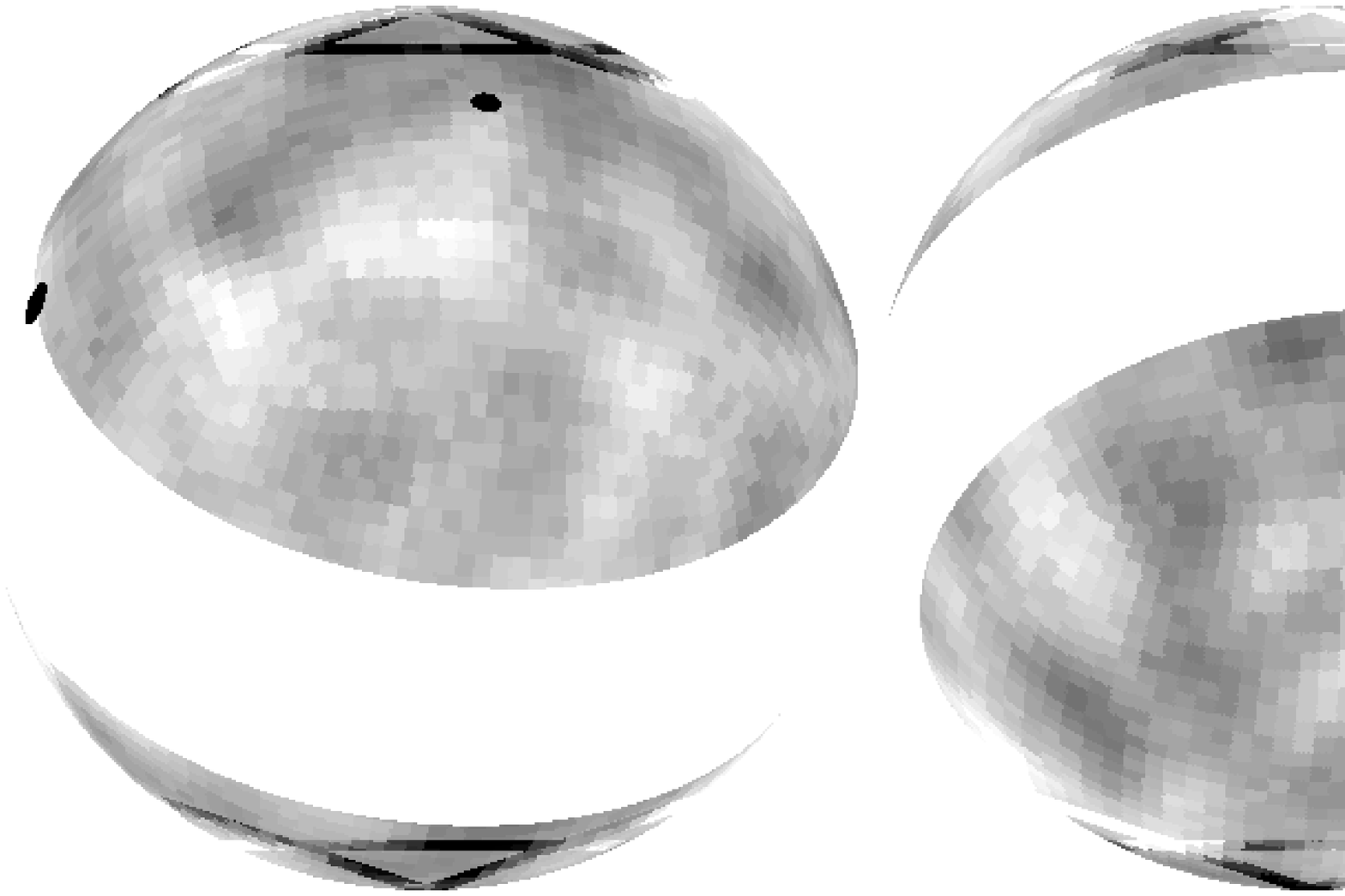"}}  } ::::
\caption[]{ \mycaptionfont
COBE DMR 10$\deg$ smoothed `combined' ASDS map of $\delta T/T$ 
in polar projection, 
in coordinates of the $T^2$ model. The $Z_{T_2}$ axis points 
`north' and `south', 
i.e. is orthogonal to the plane of the page and is located 
at the centres of the left hand and right hand images.
The $X_{T_2}$ axis is horizontal and
the $Y_{T_2}$ axis is vertical, both are in the plane of the page.
The projection is rectilinear, so that identified circles
project to straight lines.
The left hand hemisphere consists mostly of the north galactic
hemisphere, and 
black spots are marked to show the north galactic pole and a point
close to the galactic centre 
$(l^{{\sc II}}=0,b^{{\sc II}}=20\deg).$ 
Along the galactic cut ($\pm20\deg$), galactic 
longitude increases from $\approx 0\deg$ to $\approx 360\deg$
from right to left. The identified circles for 
$(i,j)=$ $(-3,9),$ $(0,9)$ and $(-3,9)$ are highlighted by multiplying
the $\delta T/T$ values by a factor of 5. 
\label{f-polar}}
\end{figure*} } 


\section{Introduction}

Since \citeauthor{Schw00}'s \citeyearpar{Schw00,Schw98}
`few remarks' which required `a total break
with the astronomers' deeply entrenched views' of zero curvature 
and trivial topology at the beginning of the twentieth century,
observational research into cosmic topology progressed very slowly
until 1993. In that year, 
a burst of articles comparing special cases or
classes of multiply connected models with 
 observations of 
the cosmic microwave background (CMB) by the COBE satellite 
\citep{Stev93,Sok93,Star93,Fang93,JFang94} were published, 
followed by the review article of \citet{LaLu95} and 
new or updated (topology independent) 
methods of analysing three-dimensional 
catalogues of conventional astrophysical
objects \citep{LLL96,Rouk96,RE97} were also invented.

Further work in analysing COBE data has been carried out, both 
for flat \citep{deOliv95,dOSS96,LevSS98a}
and for hyperbolic \citep{BPS98,Inoue99,Aurich99,CS99}
multiply connected models.

The COBE analyses simulate structure in the Universe for given
3-manifolds and estimate the probability that statistical
properties of the observed CMB temperature fluctuations could have
been drawn from distributions of those same properties for the
simulated structures. This is useful work, but requires assumptions
which are likely to be somewhat modified in the case that the Universe is
observably compact, and results in statements which may relate more to the
assumed probability distributions representing structure 
than to the statistics of measurement
uncertainty [e.g. Section 1.2 of \citet{Rouk99} or 
\SS5.3.2(ii) of \citet{LR99}].

A more direct observational approach is simply to test the
self-consistency of the CMB temperature fluctuations with the multiple
topological imaging implied by any hypothesised multiply connected
model. The exact set of points which should be multiply imaged
consists of a set of identified circles on the surface of last
scattering 
\citep{Corn96,Corn98b,Weeks98}. Given the measurement uncertainties
in the temperature fluctuations, a model can be treated 
as a null hypothesis which one tries to reject. If hypothetically
corresponding sky pixels do not have significantly different 
temperatures, then the model is consistent with the data.

This approach avoids the risk of assuming 
perturbation statistics which could be inconsistent with the hypothesis
being tested. 
An application of the identified circles principle to COBE data
has previously 
been demonstrated on an observationally motivated toroidal
universe model in \citet{Rouk99}. 

In this paper, a small value of the injectivity diameter, 
$2\rinj \equiv 2R_H/10$ (where $R_H$ is the horizon radius), 
is chosen, likely possibilities for the optimal 
orientations for a `2-torus' ($T^2 \ttimes R$ model, hereafter 
`$T^2$') are found 
based on the four year COBE DMR data,
and its consistency within measurement error of the four year
COBE DMR data is examined. To avoid confusion by non-specialist
readers, 
it is noted here that for a flat, zero cosmological
constant model, the horizon radius in comoving coordinates 
is $R_H = 2c/H_0 = 6000${\hMpc}, so that $0.8 R_H=$ $0.4 (2R_H)=$ 
$4800${\hMpc} 
and $2R_H/10 = 1200${\hMpc}. 

The constraints suggested by authors such as \citet{Stev93}, 
\citet{LevSS98a} are that $2\rinj > 8R_H/10.$ A counterexample
with $2\rinj= 2R_H/10$ is clearly sufficient to disprove the
suggested constraints.

The assumption regarding structure 
formation common to other authors' work (cited above),
i.e. that the temperature fluctuations are 
due to the na\"{\i}ve Sachs-Wolfe effect \citep{SW67},
is adopted here, but complemented by consideration of 
the integrated Sachs-Wolfe effect \citep{SW67} 
(hereafter, NSW and ISW, respectively) which the above 
authors did not quantitatively consider [though 
see, e.g. \citet{Corn98a,Corn98b}, \citet{Uzan98}, for order of
magnitude arguments regarding 
the relevance of the ISW for cosmic topology].
The ISW is expected to be significant at COBE scales for flat models
with a non-zero value of the cosmological constant 
(e.g. \citealt{WSS94,CTurok96}), though not as large as it would be
for hyperbolic models of corresponding matter densities $\Omega_0$.
Given the recent observational evidence in favour of 
a flat, cosmological constant dominated universe 
\citepf{FYTY90,FortMD97,ChY97,SCP9812}, observational motivation
for considering the ISW also exists.

The method is defined in
\SS\ref{s-method}, the observational data 
are described in \SS\ref{s-cobe}, the results are presented in 
\SS\ref{s-results} and further discussion is provided in
\SS\ref{s-conclu}.

For reviews on cosmological topology, see 
\citet{LaLu95}, \citet{Lum98}, \citet{Stark98} and \citet{LR99}. 
For workshop proceedings 
on the subject, see \citet{Stark98} and following articles,
and \citet{BR99}.
For a list and discussion of both two-dimensional and 
three-dimensional methods, 
see Table~2 of \citet{LR99} and the accompanying discussion. 
The reader should be reminded that rapid development in the 
three-dimensional methods is presently being carried out
(\citealt{LLL96,Rouk96,FagG97,RE97,RB98,Gomero99a,LLU98,FagG99a};
\citealt{ULL99,FagG99b,Gomero99b,Gomero99c}).

The Hubble constant is parametrised here 
as $h\equiv H_0/100$km~s$^{-1}$~Mpc$^{-1}.$ Comoving coordinates are
used
(i.e. `proper distances', \citealt{Wein72}, equivalent to `conformal time'
if $c=1$). Since the counterexample 
3-manifold used is $T^2,$ 
the metric assumed is flat, i.e., 
$\Omega_0 + \lambda_0 \equiv 1,$ 
where $\Omega_0$ is the present value of the density parameter 
and $\lambda_0$ is the present value of the dimensionless cosmological
constant.

\section{Method} \label{s-method}

\subsection{The Identified Circles Principle} \label{s-matchedc}

The identified circles principle was first published by 
Cornish, Spergel \& Starkman
(\citeyear{Corn96,Corn98b}), and can be briefly resumed as follows. 

The set of multiply topologically imaged points can be generated 
by considering copies of the observer in the covering space 
placed at distances
less than the horizon diameter from the observer. The 
intersection of the two last scattering surfaces (spheres) of 
the observer and a copy of the observer is a circle. But, since
the copy of the observer is physically identical to the observer,
what in the covering space appears to be two observers looking at
one circle is equivalent to one observer looking at two circles.

If the radiation from the surface of last scattering is 
isotropic, then the temperature fluctuations around 
the one circle seen by the `two observers' are identical, 
apart from measurement uncertainty, and foreground contributions
to the observed temperatures.
Hence,  the temperature fluctuations around the two circles
seen by the one observer should also be identical.

It should be noted that for this reasoning to be valid (or for the
reasoning behind perturbation assumption based methods to be valid)
for four-year COBE data, the averaging of temperature estimates over
the two transverse sizes ($\sim 1000$\hMpc) of the three-dimensional
plasma patch which generates a COBE `pixel' would need to compensate
for its thinness ($\sim 10$\hMpc) in the radial direction, since the
latter might lead to unexpected effects from the 10{\hMpc} scale. 
See fig.~14 and Section 5.3.2 of
\citet{LR99} for an illustration and brief discussion of this
question.

It is assumed here that this averaging is valid, as was
assumed implicitly by authors making calculations using perturbation
statistic based methods (e.g. \citealt{dOSS96,LevSS98a}). If it 
were {\em not} valid, then the published 
calculations claiming to show that 
$2\rinj > 8R_H/10$ for flat multiply connected universe models would
be invalid, and would be disproved
without requiring an explicit counterexample 3-manifold.
Since the gravitational potential generating the na\"{\i}ve
Sachs-Wolfe effect is expected to be reasonably smooth on
a scale of 1000{\hMpc}, it is reasonable to suppose that
the averaging process is valid, though this has not yet
been studied rigorously in the context of cosmic topology.

It should be noted, in particular, that the averaging should overcome having to
consider the Doppler contribution to the temperature fluctuations,
which is not isotropic.

\subsection{`Good' Null Hypotheses} \label{s-bestnull}

As has already been pointed out by several authors
\citep{Star93,Fang93,dOSS96}, a small 
$T^2$ universe would cause an approximate symmetry in the CMB 
if the CMB temperature fluctuations 
were only caused by the na\"{\i}ve Sachs-Wolfe effect. 
This can be thought of in terms of the identified circles by
realising that the planes of the matched circles are all 
parallel to the long $T^2$ axis. All vectors between identified
points are orthogonal to the long axis. Hence, along any circle 
on the surface of last scattering which is centred on (orthogonal
to) the long axis, many pairs of identical temperatures should exist.

A fast method of finding a `good' $T^2$ hypothesis for a given value
of $2\rinj$ is therefore to find the angular position of the long axis
which maximises the auto-correlation statistic 
$S_{T^2}(l^{{\sc II}},b^{{\sc II}}),$ 
defined
\begin{equation}
S_{T^2}(l^{{\sc II}},b^{{\sc II}}) \equiv { \left<
         { 2 \left({\delta T \over T}\right)_i 
                  \left({\delta T \over T}\right)_j }\right> 
     \over { 
\left<\left[\Delta \left({\delta T \over T}\right)\right]_i^2 +
    \left[\Delta \left({\delta T \over T}\right)\right]_j^2  \right> } } 
\label{e-st2}
\end{equation}
where $\left({\delta T \over T}\right)_i$ and $ \left({\delta T \over
T}\right)_j$ are the temperature fluctuations in two celestial
positions $(i,j)$ at equal angles from the long $T^2$ axis 
in the direction $(l^{{\sc II}},b^{{\sc II}})$, 
(i.e. along a circle orthogonal 
to the long $T^2$ axis) and
$\Delta \left({\delta T \over T}\right)_i$ and $\Delta \left({\delta T
\over T}\right)_j$ are the corresponding one standard deviation 
measurement uncertainties.
If the $T^2$ hypothesis with a 
long axis at $(l^{{\sc II}},b^{{\sc II}})$ were
correct, then some of these pairs $(i,j)$
would denote matched pixels, but many others would not. If
the Universe were simply connected, then none of the pairs would
denote matched pixels.

The highest few values of $S_{T^2}$ define the long axes of `good'
$T^2$ hypotheses. Since this symmetry statistic is only an approximate
indicator of multiple connectedness, combining matched pairs and
unmatched pairs, the short axes
need to be chosen and the identified
circles principle applied in order to see if these long axes really
do imply good multiply connected models. 
A range of orientations of the two short axes 
(assumed to be orthogonal to each other and to the long axis) needs
to be considered for long axis positions close to 
those suggested by the $S_{T^2}$ statistic.
An identified circle statistic is used to test each possibility 
(\SS\ref{s-nullhyp}). 

It is found here that this procedure 
is sufficient to find a `good' null hypothesis, 
as quantified below.

\subsection{Null Hypothesis Testing} \label{s-nullhyp}

To see if the measured temperature fluctuations are consistent with 
multiple topological imaging,
the null hypothesis that temperatures on corresponding `pixels'
are equal to within observational error is considered.

The null hypothesis is tested by considering the difference 
in the temperature fluctuations in two 
corresponding `pixels' on matched circles to be a random realisation
of a Gaussian distribution centred on zero with a width determined
by the uncertainties of the measurements in the two pixels.
By normalising the difference for each pair by the uncertainty
in that difference, 
the full set of pairs of multiply imaged
pixels is combined to form a large sample of a single
distribution $\{ d_{ij} \}$, which should have a mean of zero and a
standard deviation of unity 
if the null hypothesis is correct.

The normalised difference is defined
\begin{equation}
d_{ij} \equiv
         {  \left[ \left({\delta T \over T}\right)_i 
                 - \left({\delta T \over T}\right)_j \right]
     \over 
 \sqrt{ \left[\Delta \left({\delta T \over T}\right)\right]_i^2 +
              \left[\Delta \left({\delta T \over T}\right)\right]_j^2 } }
\label{e-dij}
\end{equation}
with the same notation as above, except that 
in this case {\em every} pair $(i,j)$ corresponds
to hypothetically identical pixels.
The standard deviation 
$\sigma$ is defined by 
 \begin{equation}
 \sigma^2(r)\equiv \left< d_{ij}^2 \right> 
 \label{e-sigma}
 \end{equation}
and the mean difference is 
 \begin{equation}
 d\equiv \left< d_{ij} \right>
 \label{e-dmean}
 \end{equation}
where $(i,j)$ vary over all pairs of points on matched circle pairs.

The observed distribution $d_{ij}$ is compared with a Gaussian distribution
of mean zero and standard deviation unity 
via a Kolmogorov-Smirnov (KS) test. Since the circles oversample 
the COBE data set, i.e. there are only $\sim 300$
independent pixels, or $\sim 150$ independent pixel pairs, 
in the COBE data set (for a $\pm20\deg$ galactic
cut and $10\deg$ resolution), a
subset of less than 
$\sim 150$ $d_{ij}$ independent values exists (since not all points 
above galactic latitudes of 20{\deg} are on identified circles).
An upper bound on the KS probability $P$ that the observed 
distribution is consistent with the null hypothesis can be provided by 
using the full set of circles but using $N=150$ in the probability 
estimate ($P_{\mbox{\rm all}}$), 
while a lower bound (and more accurate estimate) can be provided by
choosing 
an evenly spaced subset of the circles containing $\sim 150$ $d_{ij}$
values, which removes most of the correlated pixels 
($P_{\mbox{\rm subs}}$).

For completeness, the statistic of \citet{Corn98b} should be
mentioned. 
This is essentially 
a two-point autocorrelation function normalised by the variance per
pixel [eq.~(2) of \citeauthor{Corn98b}], defined
\begin{equation}
S\equiv { \left<
         { 2 \left({\delta T \over T}\right)_i 
                  \left({\delta T \over T}\right)_j }\right> 
     \over { 
\left< \left({\delta T \over T}\right)_i^2 +
    \left({\delta T \over T}\right)_j^2  \right> } } 
\label{e-corr}
\end{equation}
again using the same notation, where $i$ and $j$ 
denote matched pixels. Note that this differs from $S_{T^2}$ 
defined in \SS\ref{s-bestnull}.

\subsection{Integrated Sachs-Wolfe Effect and Other `Noise'} \label{s-isw}

Previous authors comparing flat multiply connected 3-manifolds
to statistics derived from COBE maps have ignored the ISW effect.
These foreground temperature fluctuations are generally present
on large scales, except in the special case of an 
$\Omega_0=1, \lambda_0=0$ universe.

Due to the lack of sufficient knowledge of the three-dimensional 
map of the gravitational
potential from the observer to a redshift of $z\sim 2,$ an observationally
based estimation of the ISW effect would 
require considerable model dependent extrapolation from observational
data.
However, statistical, theoretical estimates of the ISW effect 
can be made.

The observed values of $\delta T/T$ can then be treated as 
estimates of the NSW effect which contain systematic uncertainties 
of the order of magnitude of the ISW effect. The $\delta T/T$ 
contributions of the ISW effect in two multiply imaged pixels
should not be any more correlated with one another than any 
non-multiply imaged pixels. 
(`Multiply imaged' refers here only to the
surface of last scattering.)
Hence, the ISW components of the $\delta T/T$
values can be approximated as noise.

From fig.~1 of \citet{CTurok96}, it is shown that for a value of 
$\lambda_0 \approx 0.8$ (and $h=0.7$, both values of which are close
to observationally favoured values), the ISW contribution to $\delta T/T$ 
values can be nearly as much as the NSW contribution on large scales.
This can be parametrised for the present purposes by 
\begin{equation}
x^2 \equiv 
{
 \left< \left({\delta T \over T}\right)_{\mbox{\rm ISW}}^2 \right>
\over
 \left< \left({\delta T \over T}\right)_{\mbox{\rm NSW}}^2 \right>
+ \left< \left({\delta T \over T}\right)_{\mbox{\rm ISW}}^2 \right> }, 
\label{e-xdefn}
\end{equation}
so that the temperature uncertainties can be re-estimated as
\begin{equation}
\left[\Delta \left({\delta T \over T}\right)\right]^2
\equiv 
{ 
\left[\Delta \left({\delta T \over T}\right)\right]_{\mbox{\rm obs}}^2
+
x^2 \; \left< \left({\delta T \over T}\right)^2 \right> }
\label{e-isw}
\end{equation}
where 
$\left[\Delta \left({\delta T / T}\right)\right]_{\mbox{\rm obs}}^2$
is the measurement uncertainty. 

The NSW and ISW are assumed to be uncorrelated with one another, 
since gravitational structures at $z<2$ and
$z\approx 1100$ should not be significantly 
cross-correlated for a simply
connected model, and should only (at most) be very marginally 
cross-correlated in the multiply connected case.

For a simply connected model, this is simply because the distances are
very large, so cross-correlations would be very weak. 
For a multiply connected model, a single density
perturbation will in many cases make $\delta T/T$ contributions to the
ISW at differing redshifts, which, if it were somehow possible to
separate these by labelling them by their redshifts, would contribute
multiple topological `images' in differing directions and 
at differing redshifts to a CMB map, 
where an `image' is the integral over a short redshift interval
of the time varying gravitational potential in some direction. 
However, the absence
of such `redshift labels' and resulting projection
(integration) of this effect over a wide range in redshift is likely
to make both the resulting auto-correlations and the cross-correlations
with NSW contributions (at the surface of last scattering) 
small in amplitude. Hence, a conservative 
modelling of the ISW is to treat it as noise which is 
uncorrelated with the NSW.

A range of values $0.0 < x^2 < 0.6$ is considered. This can represent
the ISW effect and/or systematic uncertainties not otherwise taken
into account. Because the signal to noise ratio of the smoothed data
is $\sim 2$,
a value of $x^2=0.5$ would correspond to assuming that the
total random plus systematic uncertainties are about 
twice ($1 + 0.5 \times
2 = 2$) the random uncertainties as calculated by the COBE team.

\subsection{Observations} \label{s-cobe}

The COBE DMR four years' observational data \citep{Bennett94}
are used as recommended by the COBE team. 
These are made available 
(web address in acknowledgments)
as dipole subtracted, foreground corrected `DMR Analysed Science
Data Sets' (hereafter, ASDS). The ASDS
corrected for galactic emission by the `combination'
technique of removing synchrotron, dust and free-free emission
is used, where the weights used by the COBE team for 
combining the 31GHz, 53GHz and 90GHz maps are 
$-0.49,$ $1.42$ and $0.18$ respectively

Data between galactic latitudes
of $-20\deg$ and $+20\deg$ are not considered.

\fprobISW

Since the data set provided is oversampled, a smoothing by a 
Gaussian of $10\deg$ 
full width half maximum (FWHM) 
(differences between two beams of $7\deg$ were measured by the DMR, 
i.e. Differential Microwave Radiometer) is necessary.

\section{Results} \label{s-results}

Calculation of the $S_{T^2}$ symmetry statistic for
the ASDS map (with a 10\deg FWHM smoothing) results
in a list of `good' possibilities for the orientation of the 
long axis of a $T^2$ model. The three positions with the
highest values of $S_{T^2}$ 
are listed in Table~\ref{t-bestlong}. 
 The second and third positions are approximate antipodes, suggesting
that a `good' $T^2$ candidate should have a long axis close to 
these two positions. This indeed is the case.

\tbestlong

A search of orientations within $\sim 5\deg$ of 
these long axis positions for various possibilities of 
orthogonal short axes was performed. 
A difference set around matched circles $\{d_{ij}\}$ 
and a Kolmogorov-Smirnov comparison of  $\{d_{ij}\}$ 
with a Gaussian distribution (of mean zero and standard deviation unity)
were calculated for the different orientations.

\tbestshort

A `good' 3-manifold hypothesis found by this procedure is defined 
in Table~\ref{t-bestshort}. KS 
probabilities that the COBE data differences in supposedly
matched pixels can occur, given 
the null hypothesis that the multiple
topological imaging due to the 3-manifold chosen occurs, are 
also indicated in Table~\ref{t-bestshort} for 
the full set of circles ($P_{\mbox{\rm all}}$, which is an 
overestimate of $P$) and for a subset ($P_{\mbox{\rm subs}}$, 
which is a slight underestimate of $P$, but a better estimate than 
$P_{\mbox{\rm all}}$). An ISW/systematic error contribution 
of $x=0.3$ is assumed.

\fcircleCE 

It is clear from Table~\ref{t-bestshort} 
that this $T^2$ model is not significantly rejected,
i.e. that it is consistent with the COBE data.

\fpolar

Signficance levels at which to reject the hypothesis are shown
in Fig.~\ref{f-probISW} as a function of $x$, i.e. as a 
function of the contribution from the ISW and/or 
otherwise unaccounted for systematic errors. Since the value 
of $\sigma$ is not much greater than unity 
(e.g. $\sigma=1.6$ for $x^2=0.3$, see Table~\ref{t-bestshort}), it
is unsurprising that the hypothesis cannot be significantly rejected.

Fig.~\ref{f-probISW} shows that even if the COBE team's estimates
of the uncertainties are taken to include all random and systematic
error and the ISW is ignored, i.e. $x^2=0,$ the model is still rejected
at only 92\%, i.e. less than what is considered a high significance level.


What are the actual values of the temperature fluctuations along the
identified circles? The identified circles for the $T^2$
model are shown in Fig.~\ref{f-circleCE}, for $x^2=0.3.$
The values and overall features of $\delta T/T$ 
along the matched circles are not, in general, significantly different, 
apart from one section of the panel for $(i,j)=$ $(3,0).$

Fig.~\ref{f-polar} shows the COBE map in polar projection, 
in the $T^2$ model coordinates, with some examples of the matched
circles indicated. As explained in \SS\ref{s-bestnull}, 
an approximate circular symmetry can be expected around the 
long axis of a $T^2$ model. 
Visual inspection of Fig.~\ref{f-polar} suggests that 
the `northern' hemisphere does have some circular symmetry,
though circular symmetry is less obvious in the `southern' hemisphere.

\section{Discussion and Conclusions} \label{s-conclu}

Whereas flat multiply connected models were supposed to be excluded by
COBE data for $2\rinj \le 0.4(2R_H),$ a simple flat $T^2$ model of
$2\rinj = 0.1(2R_H)$ with long and short axes at $(l^{{\sc
II}},b^{{\sc II}}) =$ $(280\deg,37.5\deg),$ $(184\deg,8\deg)$ and
$(264\deg, -51\deg)$ respectively has been shown to be consistent with
the COBE four year data.

This result does not show that {\em calculations} by previous authors
were incorrect, but instead shows that some of the conclusions drawn
from the calculations were over-extrapolated and somewhat overstated,
with the risk of misinterpretation by non-specialist readers.

What are the differences between this result and previous
authors' work? The primary difference is the use of the 
identified circles principle rather than the use of indirect
statistics. 

Another difference, made possible by the use of the identified circles
principle, is that direct observational consistency
between hypothesis and data is tested,
and reliance on simulations and
assumptions on the statistics of density perturbations is avoided. 

If one tried to reconstruct the three-dimensional field 
of density fluctuations in the fundamental polyhedron of
the $T^2$ model found here, it could be the case that
their statistical properties are not quite Gaussian.
However, this implication would {\em not} violate 
the majority of COBE analyses showing that the COBE data are
consistent with Gaussianity, since,
by construction, such a three-dimensional
field would be directly consistent with the COBE map itself.
[The same argument applies for the results of 
authors who find evidence of 
non-Gaussianity \citep{Ferr98,Pando98}.]

In other words, if one calculated a statistical analysis of the
temperature fluctuations corresponding to the subset of the
reconstructed three-dimensional perturbation field which is restricted
to the surface of last scattering, then this statistical analysis
would be identical (within statistical error) to corresponding
published analyses of COBE four-year data. The $C_l$
(spherical harmonic) spectrum of the $T^2$ solution found here
is that of fig.~4 of \citet{Gor96} and fig.~1 of \citet{Teg96}.
It is consistent with the $C_l$ spectrum of the COBE data, since
it {\em is} the spectrum of the COBE data.

A third difference is that the ISW is treated here (in a simple but
quantitative way).  However, Fig.~\ref{f-probISW} shows that 
it is not necessary for the ISW to be dominant (greater in amplitude
than the NSW component) for the 3-manifold to be consistent with
the COBE data. An ISW component from half to equal the amplitude of 
the NSW component, i.e. where $0.33 \le x^2 \le 0.5$, implies 
$80\% \ge 1-P \ge 50\%$ rejection levels 
of the 3-manifold respectively, i.e. implies that 
the 3-manifold is consistent with the COBE data even though the ISW
component is weak. Inclusion of an 
ISW component in the perturbation statistic methods cited above
would modify the results of those calculations, but is unlikely that
it would invalidate them.

It could also be argued that there is a philosophical 
difference relative to the simulation based results.
This is expressed statistically and is 
sometimes titled the `cosmic variance' argument.

The analyses using simulations quote statistical confidence levels
relating to an ensemble of `observable' universes 
(where an observable universe is equivalent to 
the space up to our own horizon), 
either theoretical observable universes or different samples 
of the one physical Universe which are observable in principle 
if one waits many Hubble times. It is then assumed by the 
Copernican principle that parameters of our observable Universe 
should be within a few standard deviations of the mean of any measurable 
parameter.

However, if the Universe is observably small, then there exist
no other samples of observable universes, and the theoretical
properties are unlikely to be exactly identical to those generally
assumed. 

\tworst

On the other hand, the confidence levels in this study relate only to
measurement uncertainties in the $\delta T/T$ values, plus extra
allowance for the integrated Sachs-Wolfe effect (for a high value of
the cosmological constant $\lambda_0=1-\Omega_0$) and possible other
systematic measurement uncertainties. The cosmic variance 
argument is not invoked.

Although it has been shown that a $2\rinj= 2R_H/10$ multiply 
connected model of the Universe which is consistent with the 
COBE four year data can easily be found, it is {\em not} the theme of
this paper to claim that this should be considered a strong
candidate for the 3-manifold of the Universe. Finding one
counterexample is not the same task as finding the best
counterexample, nor is it anywhere near sufficient to
reject the simple connectedness hypothesis.
For example, as in the case of other authors, $T^2$ models
are only considered for which the faces of the fundamental 
polyhedron are mutually orthogonal. 

Nevertheless, the $T^2$ model found could be considered
as a useful working hypothesis, or as a candidate 3-manifold, for
use in working carefully through the details of methods, 
analyses or searches for systematic errors, both for 
two-dimensional (CMB) and three-dimensional [catalogues of
collapsed luminous objects \citep{LLL96,Rouk96,RE97}] studies.

The reader should be reminded that application of the identified
circles principle is powerful enough to reject small universe
hypotheses with the COBE four-year observational data, even in the
presence of a strong ISW effect. 

For example, use of the $S_{T^2}$
test as above (\SS\ref{s-bestnull}) to find examples of `bad' $T^2$
hypotheses (by the most negative values of $S_{T^2}$) 
and followup by null hypothesis testing (as in \SS\ref{s-nullhyp}) 
leads to the results in Table~\ref{t-worst}. If the
$T^2$ model defined in Table~\ref{t-worst} 
were correct, then even with $x^2=0.6$ 
[eq.~(\ref{e-xdefn})], 
i.e. with a large contribution of the ISW effect which hides a lot
of the signal from the surface of last scattering, 
then the probability of obtaining the COBE observations is
$P(\mbox{\rm subs}) = 0.001
\ltapprox P < P(\mbox{\rm all}) = 0.002$. In other words,
the hypothesis is rejected at the $99.8\% < 1-P $ level.

Future uses of 
the present study may help to suggest ways of searching for
systematic errors in CMB data, and trying to find ways of avoiding
these.

For example, 
the seriousness of contamination by the galactic plane for 
cosmic topology studies is clear in Fig.~\ref{f-polar}. A large
fraction of pixel pairs which could contribute to strengthening or 
weakening of a $T^2$ hypothesis are invalid because one or the
other of the pixels in the pair lies close to the galactic plane.

Other topologically non-trivial 3-manifolds could obviously have
similar problems.

Another possibility would be to follow up the circular symmetry 
evident in Fig.~\ref{f-polar}. 
If there were a valid
physical argument explaining why the northern galactic hemisphere
COBE measurements had a circular symmetry imposed, then that
could refute this particular counterexample. This would imply that 
another would have to be sought based on the new effect, but would
also imply an effect which would need to be taken into account
for analyses of MAP and Planck Surveyor data --- whether for cosmic 
topology or other purposes.

Alternatively, if there were a valid
physical argument explaining why the southern galactic hemisphere
COBE map (but not the northern hemisphere) 
contained a systematic error of the order of magnitude
of the measurement error already present, then the specific 
$T^2$ model found here 
would be considerably strengthened.

In the absence of such a physical argument, 
the $T^2$ model found simply remains
a counterexample to the claimed `constraints'.

\section*{Acknowledgments}

Helpful discussions and encouragement from 
Ubi Wichoski, Suketu Bhavsar and Tarun Souradeep 
were greatly appreciated. 
Use was made of the COBE datasets 
{\em (http:// www.gsfc.nasa.gov/ astro/cobe/ cobe\_home.html)}
which were developed by NASA's Goddard Space Flight Center 
under the guidance of the COBE Science Working
Group and were provided by the NSSDC. 

\subm \clearpage ::::

\end{document}